\documentclass[twocolumn,
superscript address]{revtex4}

\usepackage{fancyhdr,amsmath,amscd,amssymb}
\usepackage{amssymb,epsf,graphicx}
\usepackage{amsmath}
\usepackage{booktabs}
\usepackage{epsfig}

\usepackage{epsfig}
\usepackage{latexsym}
\usepackage{graphicx}
\usepackage[T1]{fontenc}
\usepackage[latin9]{inputenc}
\usepackage{array}
\usepackage{color}

\usepackage{calc}
\usepackage{tikz}
\usepackage{slashbox}

\begin{document}

\title{Effect of the Minimal Length on Bose-Einstein
Condensation in the Relativistic Ideal Bose Gas}

\author{Xiuming Zhang}
\thanks{Corresponding author. zhangxm@uestc.edu.cn}
\affiliation {School of Physical Electronics, University of
Electronic Science and Technology of China, Chengdu, 610054, P. R.
China}

\author{Chi Tian}\thanks{rectaflex@gmail.com}
\affiliation {School of Physical Electronics, University of
Electronic Science and Technology of China, Chengdu, 610054, P. R.
China}


\begin{abstract}
Based on the generalized uncertainty principle (GUP), the critical temperature and the Helmholtz free energy of Bose-Einstein condensation (BEC) in the relativistic
ideal Bose gas are investigated.
At the non-relativistic limit and the ultra-relativistic limit, we calculate the analytical form of the shifts of the critical temperature and the Helmholtz free energy caused by weak quantum gravitational effects. The exact numerical results of these shifts are obtained.
Quantum gravity effects lift the critical temperature of BEC.
By measuring the shift of the critical temperature, we can constrain the deformation parameter $\beta_0$. Furthermore, at lower densities, omitting quantum gravitational effects may lead to a metastable state while at sufficiently high densities, quantum gravitational effects tend to make BEC unstable. Using the numerical methods, the stable-unstable transition temperature is found. 
\end{abstract}

\maketitle

In the absence of a full theory of quantum gravity, effective models are useful tools to gain some experimental signatures from quantum theory of gravity.
This is so-called quantum-gravity phenomenology or Planck-scale phenomenology \cite{Amelino-Camelia2002MPLA,Amelino-Camelia2013LRR}.
Most studies focused on the gamma-ray astrophysics \cite{Arkani-Hamed1999PRD,Amelino-Camelia1998nature393,Biler1999PRL83,Kifune1999APJL},
fundamental particle processes \cite{Protheroe2000PLB,Amelino-Camelia2001PRD,TJacobson2002PRD,Amelino-Camelia2002PLB},
neutrino physics \cite{Alfaro2000PRL,Jacob2007NaturePhysics} and the laser-interference of gravity waves \cite{Amelino-Camelia1999Nature398,YJNg2000FP,Amelino-Camelia2000PRD62,Amelino-Camelia2004CQG,Hogan2012PRD},
where particles exist in the ultra-relativistic regime. %
It is also possible to test quantum gravity effects using cold or slow atoms, where particles exist in the non-relativistic regime \cite{Amelino-Camelia2009PRL,Mercati2010CQG,BrisceseGrether2012EPL,Briscese2012PLB}.  In \cite{BrisceseGrether2012EPL,Briscese2012PLB,Castellanos2013EPL,Castellanos2012MPLA,Castellanos2014PLB}, quantum gravity effects of non-traped and harmonically trapped Bose-Einstein condensates are examined respectively using the deformed non-relativistic free-particle  energy-momentum dispersion relation.
Quantum gravity effects cause an explicit shift in the condensation temperature, and then ultra-precise measurements of the condensation temperature make it possible to upper-bound the deformation parameter. Therefore, BEC opens a new road to quantum-gravity phenomenology.

The effective quantum gravity models used in the above are modified dispersion relations (MDR). Quantum gravity theories also predict that gravity itself leads to an effective cutoff in the ultraviolet, i.e., a minimal observable length \cite{Garay1995IJMPA,Scardigli1999PLB}.
Some realizations of the minimal length from various scenarios are proposed. One of the most important models is the generalized uncertainty principle (GUP), derived from the generalized
commutation relation
\begin{equation}
\left[x,p\right]=i\hbar\left(1+\beta p^2\right),\label{simpleparameterGUP}
\end{equation}
where $\beta=\beta_0 l_p^2/\hbar^2 =\beta_0/c^2 M_p^2$ with the Planck
mass $M_p=\sqrt{\hbar c/G}$ and the Planck length
$l_p=\sqrt{G\hbar/c^3}$. $\beta_0$ is a dimensionless parameter.
And then, we can get the familiar form of GUP
\begin{equation}
\Delta x\Delta p\geq \frac \hbar 2 [1+\beta (\Delta p)^2],\label{GCR2}
\end{equation}
this in turn gives the minimum measurable length
\begin{equation}
\Delta x\geq \Delta_{min}\equiv\hbar \sqrt{\beta}=\sqrt{\beta_0} \;l_p.\label{minimallength1}
\end{equation}

In this letter, based on the generalized uncertainty principle (GUP), we are going to investigate the effects of the minimal length on Bose-Einstein
condensation in the relativistic ideal Bose gas. Instead of considering thermodynamical functions of the relativistic ideal Bose gas,
we focus on the BEC critical temperature and the stability of relativistic ideal Bose-Einstein condensates in the presence of quantum gravity.
We will compare the exact Helmholtz free energy
between ordinary particle-antiparticle system and the system including quantum gravity at all number densities and all critical temperatures. It will be shown that when the critical temperature $T_c$ is low enough, the Helmholtz free energy with quantum gravity becomes lower, thus implying that the ideal Bose gas considering quantum gravity is stabler. However, when the critical temperature $T_c$ is high enough, the Helmholtz free energy with quantum gravity becomes higher. This implies that quantum gravity effects make the Bose-Einstein
condensation of ideal Bose gas unstabler. Using numerical methods, we can find the stable-unstable transition temperature.

We consider a gaseous system of noninteracting bosons and antibosons of mass $m$ whose macro state is denoted by $(T, V, \mu)$,
where $\mu$ is the chemical potential of
bosons confined in volume $V$ at a temperature $T$. When considering the effects of minimal length, in three dimensional space,
the revised state density function has been derived from Eq.\,(\ref{simpleparameterGUP}) , which is given \cite{Kempf1995PRD,LNCMinic2002PRD,Fityo2008PLA}
\begin{equation}
D(p)dp=\frac{ V p^2}{2\pi^2\hbar^3 \left(1+\beta p^2\right)^3} dp,\;\; p\in (0,+\infty).\label{sdfqgs}
\end{equation}
This implies that, compared with the ideal gas without quantum gravity, the ideal gas in the GUP scenario has a lower probability for the particles to occupy the same quantum state.
The revised state density function reaches a maximum around $p=1/\sqrt{2\beta}$.
Using the revised state density (\ref{sdfqgs}), the logarithm of the grand partition function for this system can be written in integrated form with
relativistic dispersion relation $E_p=\sqrt{m^2
c^4+p^2c^2}$ as
\begin{eqnarray}
\ln{Z_G} 
&=&\ln{Z_G}|_{p=0}\;-\frac{ V}{2 \pi^2 \hbar^3} \int_{0^+}^\infty
\frac{p^2}{\left(1+\beta p^2\right)^3} dp\times\nonumber\\
&&\left\{\ln{\left[1-\exp{\left[-\beta_\textrm{B}\left(m^2
c^4+p^2c^2\right)^{1/2}-\alpha\right]}\right]}\right. \nonumber\\
&&+\left.(\alpha \rightarrow
-\alpha)\right\},\label{grandpartitionfunction12}
\end{eqnarray}
where $\beta_\textrm{B}=1/k_\textrm{B}T$, $k_\textrm{B}$ is the Boltzmann constant and $\alpha\equiv -\mu/k_\textrm{B} T$. All other thermodynamic functions can be derived from $\textrm{ln}Z_G$. In particular,
the complete number density of boson and antiboson $n$ is
\begin{eqnarray}
n&\equiv &n_b-n_{\bar{b}}=n|_{p=0}+n|_{p>0}\nonumber\\
&=&n|_{p=0}+ \frac{1}{2\pi^2 \hbar^3} \int_{0^+}^\infty \frac{p^2}{(1+\beta p^2)^3} dp \nonumber\\
&\times &\left\{ \frac{1}{\exp{\left[\frac{\left(\sqrt{m^2c^4+p^2c^2}-\mu\right)}{k_\textrm{B}T} \right]}-1}-(\mu\rightarrow -\mu) \right\},
\label{completenumberdensitygeneral}
\end{eqnarray}
the exact Helmholtz free energy per unit volume
\begin{equation}
f\equiv\frac FV=n m c^2 -k_\textrm{B} T \frac{1}{V}\ln{Z_G} .\label{freeenergydensity0}
\end{equation}

Since $n_b$, $n_{\bar{b}}>0$ for all $p\geq 0$ and $E_0=mc^2$, the chemical potential must be bounded by $\mid \mu\mid \leq mc^2$.
We impose the constancy of $n$ to extract the correct BEC  critical transition temperature $T_c$. At the BEC critical temperature $T_c$, $ n|_{p=0}=0$, $\mid\mu(T_c)\mid=mc^2$, we set
\begin{equation}
\tilde n\equiv n\Delta_{C}^3=\left(\frac{\Delta_C}{\bar d}\right)^3,\;x\equiv\frac{pc}{k_\textrm{B}T_c},\;\lambda\equiv\frac{k_\textrm{B}T_c}{mc^2},
\end{equation}
\begin{equation}
\sigma\equiv \frac {m c^2}{E_\textrm{H}}=\sqrt{\beta_0}\,\frac{m}{M_p}=
\frac{\Delta_{min}}{\Delta_C},\;
\kappa\equiv \lambda\sigma=\frac{T_c}{T_\textrm{H}}
,
\end{equation}
\begin{equation}
F\equiv\tilde{F} mc^2,\;V\equiv\tilde{V} \Delta_C^3, f\equiv \frac FV\equiv\tilde f \,mc^2 \Delta_C^{-3},
\end{equation}
where $\bar{d}\equiv n^{-1/3}$ is the average distance between particles, $\Delta_C=\hbar/m c$ is the Compton wavelength, $E_\textrm{H}\equiv c/\sqrt{\beta}=M_pc^2/\sqrt{\beta_0}$ is the Hagedorn energy and $T_\textrm{H}=E_\textrm{H}/k_\textrm{B}$ is the Hagedorn temperature defined in \cite{Xiumzhang2010JHEP}. Then, at the BEC critical temperature, Eq. (\ref{completenumberdensitygeneral})
can be written
in dimensionless form as
\begin{eqnarray}
\tilde{n}&=& \frac{\lambda^3}{2\pi^2} \int_{0^+}^\infty \frac {x^2\;dx}{(1+
\kappa^2x^2)^3} \nonumber \\
&\times &\left\{\frac{1}{\exp{\left[\left(x^2+\frac 1{\lambda^2}\right)^{1/2}-\frac 1{\lambda}\right]}-1}-(\lambda\rightarrow -\lambda)\right\}.\label{completenumberdensityBEC12}
\end{eqnarray}
Eq. (\ref{completenumberdensityBEC12}) gives the implicit expression for $\lambda $ as a function of $\tilde{n}$ and $\sigma$, $\lambda=\lambda(\tilde{n},\sigma)$.
Furthermore, the dimensionless form of the Helmholtz free energy
\begin{eqnarray}
&&\tilde{f}=\tilde{n} +\frac{\lambda^4}{2\pi^2}  \int_{0^+}^\infty \frac {x^2\;dx}{(1+\kappa^2 x^2)^3} \times\;\;\;\;\;\;\;\;\;\;\;\;\;\;\;\;\;\;\;\;\;\;\;\;\;\;\;
\;\; \;\;\;\;\;\nonumber \\
&&\left\{ \ln{\left[1-\exp{\left[\frac{1}{\lambda}-\left(x^2+\frac {1}{\lambda^2}\right)^{\frac 12}\right]}\right]} +(\lambda\rightarrow -\lambda) \right\}
.\label{Hfreeenergy1}
\end{eqnarray}

There are different upper bounds about $\beta_0$
from different theories.  A better bound is gained from simple electro-weak consideration $\beta_0< 10^{34}$. The observed masses of neutron stars ($\leq 2 M_\odot$) indicate that $\beta_0< 10^{37}$ \cite{Xiumzhang2012PLB}. A relatively rough but stronger restriction is estimated in \cite{BrauBuisseret2006PRD}. In \cite{DasVagenas2008PRL}, based on the precision measurement of Lamb shift, an upper bound of
$\beta_0$ is given by $\beta_0< 10^{36}$. On the other hand, the values of the boson mass $m$ in cosmological observations vary from very heavy values of the order of $10^{13}$ GeV to values as light as $10^{-23}$ eV \cite{LAULopez2009JCAP}. Then we have
$
\sigma=10^{-51}\sqrt{\beta_0}\sim 10^{-6}\sqrt{\beta_0}.
$
Note that $\sigma=\Delta_{min}/\Delta_{C}$, physically, we constrain $\sigma<1$.

The size of $\kappa$ tells when the BEC critical temperature enters the level where quantum gravity effects play an important role. Naturally, for $m\neq 0$, $\Delta_{min}=0$ means that $\kappa=0$. %
For $\kappa\ll 1$, from Eq. (\ref{completenumberdensityBEC12}) and (\ref{Hfreeenergy1}), we can expand the denominator of the integrand
before performing integration and then acquire small corrections of the
critical temperature and the Helmholtz free energy due to quantum gravity effects.
When $\kappa$ approaching $1$, the critical temperature enters the level which is dominated by quantum gravity.
Although the GUP is an effective model, we will explore some properties of this region using numerical methods.


Now we consider the shift in the critical temperature. For $\kappa\ll 1$, we address two extreme cases as follows.

(i) \textit{Non-relativistic limit}, i.e., $\lambda \ll 1 $. In this situation, $e^{-1/\lambda}\simeq 0$, $[x^2+(1/\lambda)^2]^{1/2}-1/\lambda\simeq \lambda x^2/2 $.
Eq.\;(\ref{completenumberdensityBEC12}) reduces to the following integrand
\begin{eqnarray}
\tilde{n}&\simeq & \frac{\lambda^3}{2\pi^2} \int_0^\infty x^2(1-3\kappa^2 x^2)\,\frac 1{e^{\frac 12\lambda x^2}-1}dx \nonumber\\
&=& \left(\frac \lambda{2\pi}\right)^{3/2}\zeta(3/2)\left[1-9\kappa\sigma\frac{\zeta(5/2)}{\zeta(3/2)}\right],
\end{eqnarray}
where $\zeta (s)$ is the Riemann zeta function. Since $\lambda=\lambda(\tilde{n},\sigma)$, then the shift in the condensation temperature due to the quantum gravity effects is
\begin{eqnarray}
\frac{\Delta\lambda}{\lambda} &\equiv& \frac{\lambda(\tilde{n},\sigma)-\lambda(\tilde{n},0)}{\lambda(\tilde{n},0)}\nonumber\\
&\simeq &6\kappa\sigma\frac{\zeta(5/2)}{\zeta(3/2)}\simeq  \frac{12\pi\zeta(5/2)}{[\zeta(3/2)]^{5/3}}\,\sigma^2\tilde{n}^{2/3}\label{NRTEmperaturedelata11}\\
&\simeq &10.21\,\beta_0\left(\frac{\hbar}{M_pc}\right)^2n^{\frac 23}=10.21 \left(\frac{\Delta_{min}}{\bar{d}}\right)^2.\label{NRTEmperaturedelata}
\end{eqnarray}
The shift is independent on boson mass $m$ and is a monotonically  increasing function of $n^{2/3}$.
In the ${}^{85}_{37}\textrm{Rb}$ BEC with number density $n\simeq 10^{18} \textrm{m}^{-3}$ and
boson mass $m\simeq 150\times 10^{-27} \textrm{kg}$ \cite{Grether2008IJMPB}, $\Delta\lambda/\lambda\sim 10^{-58}\beta_0$.
These results are explicitly different from results obtained in \cite{BrisceseGrether2012EPL}, where $\Delta T_c/T_c^0\propto n^{-1/3}$.
In our scenario, the effects of minimal length play an important role only in the high densities or temperatures, and then thermodynamics should be corrected only in the high densities or temperatures.

(ii) \textit{Ultra-relativistic limit}, i.e., $\lambda \gg 1 $.
In this situation, $e^{1/\lambda}\simeq 1+1/\lambda$, $[x^2+(1/\lambda)^2]^{1/2}\simeq x $. $\lambda \gg 1$ implies 
$\Delta_{min}\ll \Delta_C$, $\bar d\ll \Delta_C$. From Eq. (\ref{completenumberdensityBEC12}), we obtain
\begin{eqnarray}
\tilde{n}&\simeq&\frac{\lambda^2}{\pi^2} \int_0^\infty x^2(1-3\kappa^2 x^2)\frac{e^x}{(e^x-1)^2}dx\nonumber\\
&=&\left(\frac{1}{3}-\frac 45\pi^2\kappa^2\right) \lambda^2.\label{completenumberdensityBEC122}
\end{eqnarray}
Then the shift in the
condensation temperature is
\begin{eqnarray}
\frac{\Delta\lambda}{\lambda}&\simeq &\frac {6}{5}\pi^2\kappa^2 \simeq\frac {18}{5}\pi^2\sigma^2\tilde n\label{ultrarelambda11}\\
&=&\frac {18}{5}\pi^2\,\beta_0\frac{\hbar^3}{M_p^2c^3m}n
=\frac {18}{5}\pi^2\frac{\Delta_C}{\bar d}\,\left(\frac{\Delta_{min}}{\bar{d}}\right)^2.\label{ultrarelambda}
\end{eqnarray}
The shift is dependent on boson mass $m$ and is a monotonically increasing function of $n$.
For the mass of neutron $m_n=1.6749\times 10^{-27}\,\textrm{kg}$ and the normal nuclear density $\rho= 2.7\times 10^{17}\,\textrm{kgm}^{-3}$,
$\Delta\lambda/\lambda\sim 10^{-40}\beta_0$. For the upper limit of observed values, $\beta_0=10^{36}$, the shift in the condensation temperature is $10^{-4}$,
which is far below the current laboratory measurement accuracy, $10^{-2}$. In the BEC stars, this may have little effects. However, such a small shift might produce a significant impact on the structure formation in the early universe.

\begin{figure}[h]
\centering \includegraphics[height=5cm]{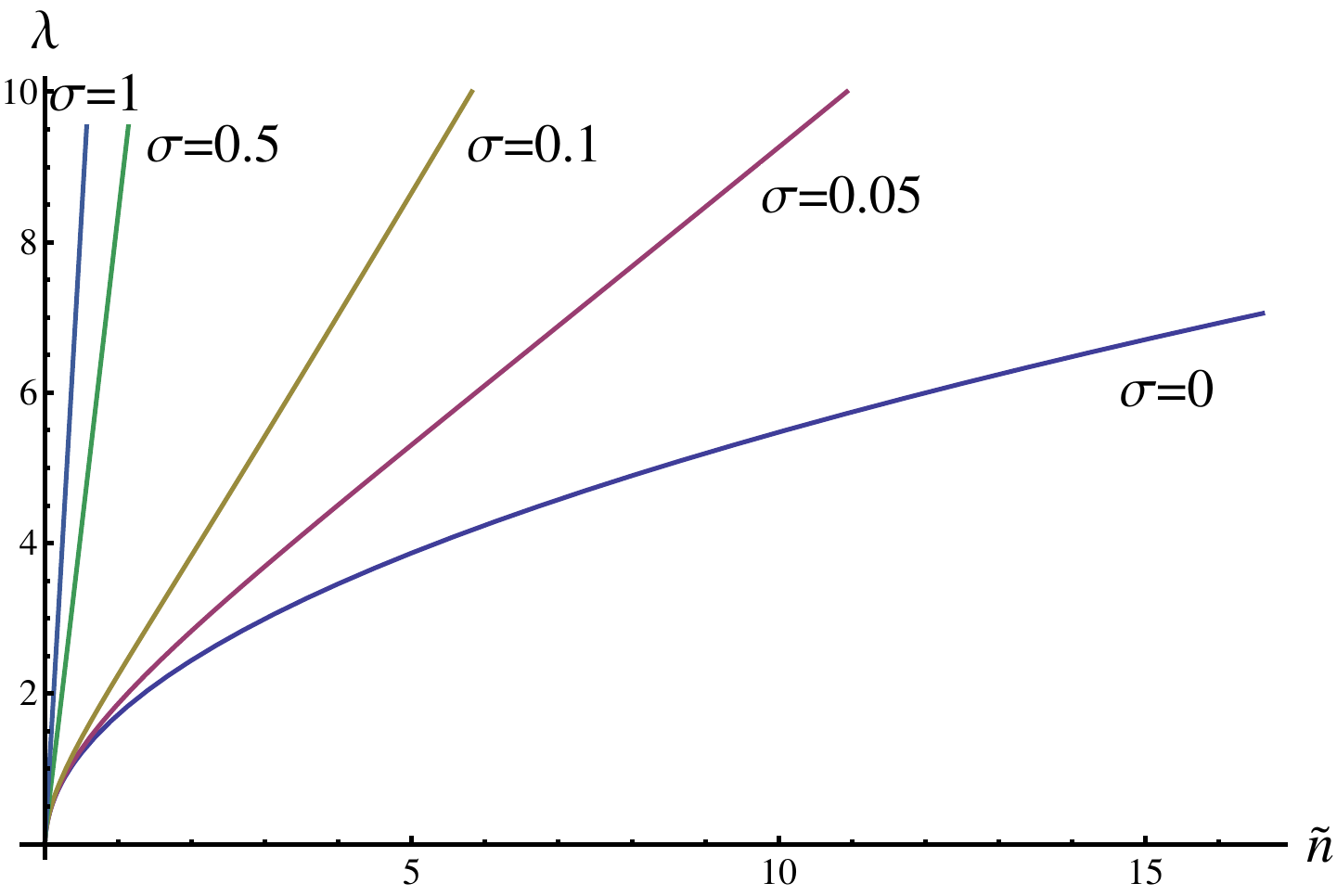} \caption {The number density $\tilde{n}$ versus the critical temperature $\lambda $ for $\sigma=0, 0.05,
0.1, 0.5,1$. The curves of $\sigma> 0$ are in the above of the curve $\sigma=0$. This means that quantum gravity effects lift the critical
temperature of BEC.}
.\label{lambdaVSn}
\end{figure}

\begin{figure}[h]
\centering \includegraphics[height=5cm]{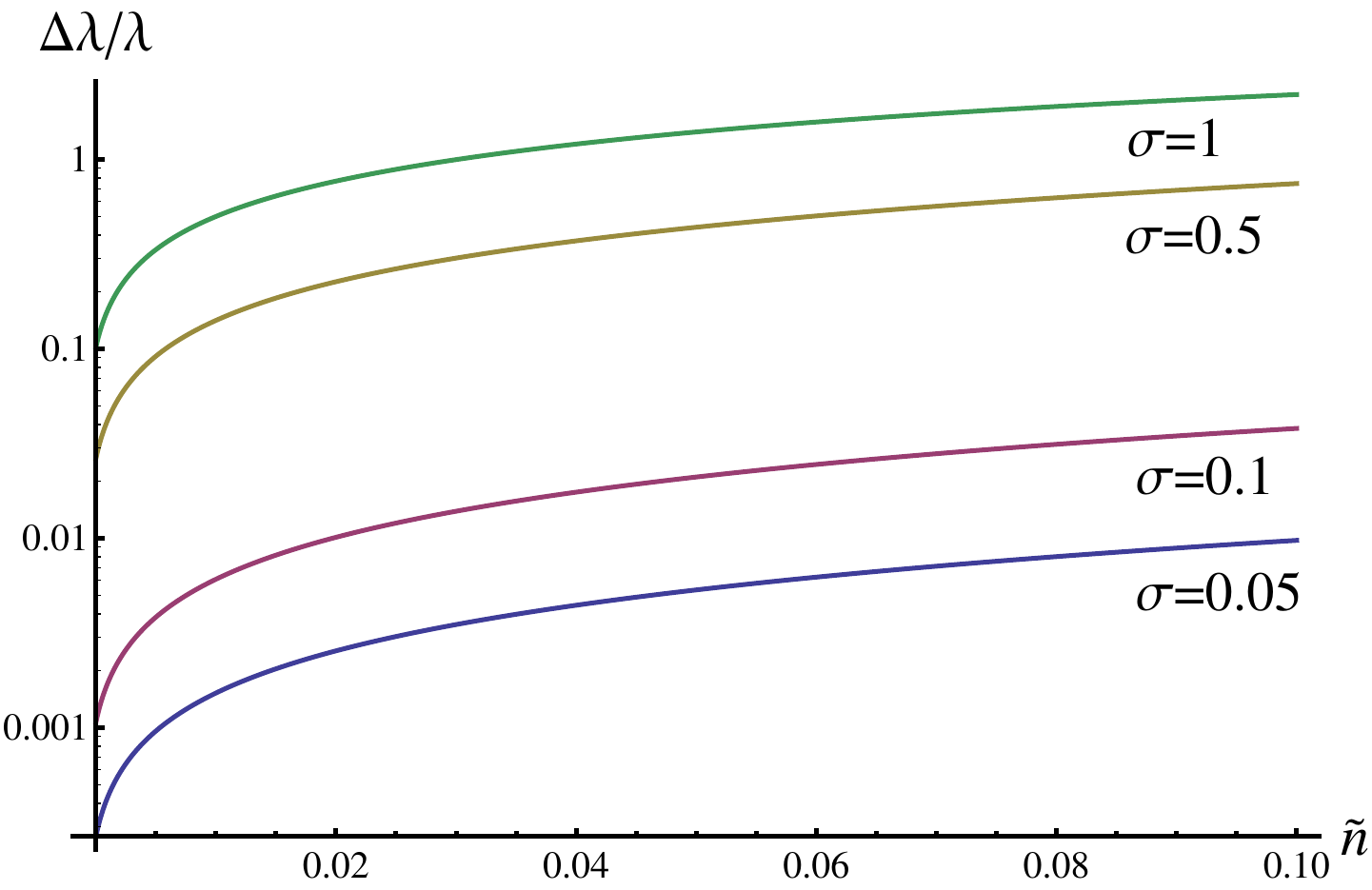}  \caption {The shift of the critical temperature ${\Delta\lambda}/{\lambda} $ versus the number
density $\tilde{n}$ for $\sigma=0.05,
0.1, 0.5,1$. In order to display on a graph for different values of $\sigma$, ${\Delta\lambda}/{\lambda} $ is displayed in the form of a logarithmic graph.}\label{deltalambdaVSn}
\end{figure}

As functions of the dimensionless boson number density $n$ (in units of $1/\Delta_{C}^3 $), FIG. \ref{lambdaVSn} displays the behavior of the critical temperature $T_c$ (in units of $mc^2/k_\textrm{B}$) numerically extracted from (\ref{completenumberdensityBEC12}) for $\sigma=0,0.05,0.1,0.5,1$.
At the low densities ($\tilde n\ll 1$), quantum gravity effects give small corrections for all $\sigma$.
At the high densities ($\tilde n\gg 1$), when $\sigma$ is close to zero, the corrections are still small since $\kappa =\lambda\sigma\rightarrow 0$.
These results are consistent with the results given by the analytical solution (\ref{NRTEmperaturedelata11}) and (\ref{ultrarelambda11}).
FIG. \ref{deltalambdaVSn} depicts the behavior of the shift of the critical temperature for $\sigma=0.05,0.1,0.5,1$.
It is easy to find that quantum gravity effects lift the critical temperature of BEC and larger $\beta_0 $ leads to more obvious lift of the critical temperature. Therefore, quantum gravity effects enhance the formation of BEC, that is, at the same critical temperature, BEC can also occur even if at relatively low density.

The shift of the critical temperature ${\Delta\lambda}/{\lambda} $ is a reflection of minimal length.
Although the interboson interactions will result in the shift, we still expect to measure and constrain the deformation parameter $\beta_0$ by measuring and constraining the shift of the critical temperature
in the relativistic ideal Bose gas.
Numerically extracted from (\ref{completenumberdensityBEC12}), Table 1 shows the values of $\sigma$ for different number
density $\tilde n$ and the shift of the critical temperature ${\Delta\lambda}/{\lambda} $.
For the mass of neutron $m_n=1.6749\times 10^{-27}\,\textrm{kg}$ and the normal nuclear density $\rho= 2.7\times 10^{17}\,\textrm{kgm}^{-3}$, $\tilde n\sim 10^{-3}$, $\Delta_C=0.1\bar d$.
From Table 1 and FIG. 2, for $\tilde n = 10^{-3}$, if we constrain  ${\Delta\lambda}/{\lambda}< 10^{-2}$, we have $\sigma =\sqrt{\beta_0}\,{m_n}/{M_p}  < 0.31$.
Then we can constrain $\beta_0<1.6\times 10^{37}$, compatible with that from the observed masses of neutron stars \cite{Xiumzhang2012PLB}. But this is not as good as the upper bound $\beta_0<10^{36}$ from the precision measurement of Lamb shift \cite{DasVagenas2008PRL}.
In our scenario, higher density and smaller temperature shift will give a more stronger upper bound.

{\small{\textbf{Table 1}.\;$\sigma$ for different number
density $\tilde n$ and the shift of the critical temperature ${\Delta\lambda}/{\lambda} $.}}
\begin{center}
\begin{tabular}{|l|*{8}{c|}}\hline
\backslashbox{$\Delta\lambda/\lambda$}{$\tilde n$}
        &$10^{-6}$      & $10^{-3}$  &1          &$10^3$    \\\hline
0.1     &10.68         &1.05       &$5.81   \times 10^{-2}  $  &$1.85\times 10^{-3}$   \\\hline
0.01    &3.15          &0.31       &$1.69   \times 10^{-2}  $  &$5.36 \times 10^{-4} $  \\\hline
0.001   &0.99          &0.10       &$5.28    \times 10^{-3}  $  &$1.68 \times 10^{-4}$  \\\hline
0.0001  &0.31          &0.03       &$1.67    \times 10^{-3}  $  &$5.31 \times 10^{-5}$  \\\hline
0.00001  &0.10          &0.01       &$5.27    \times 10^{-4}  $  &$1.68 \times 10^{-5}$  \\\hline
\end{tabular}
\end{center}

It is then tempting to speculate that quantum gravity will affect the stability of BEC.
For $\kappa\ll 1$, we also address two extreme situations. At the non-relativistic limit, the Helmholtz free energy
\begin{eqnarray}
\tilde{f}&\simeq&\tilde{n} +\frac{\lambda^4}{2\pi^2}  \int_{0^+}^\infty  {x^2}{(1-3\kappa^2 x^2)}\ln{\left[1-e^{-\frac 12 \lambda x^2}\right]dx} \nonumber \\
&\simeq & \left(\frac \lambda{2\pi}\right)^{3/2}\left[\zeta(\frac 32)-\zeta(\frac 52)\lambda\right]-\frac {18\pi\,\zeta(5/2)}{[\zeta(3/2)]^{5/3}}\,\sigma^2\tilde n^{5/3}\nonumber\\
&\equiv & \tilde f^{NR}_0+\Delta f^{NR}(\tilde n,\sigma),\label{kappax1freeenr1}
\end{eqnarray}
where $\tilde f^{NR}_0=(\lambda/{2\pi})^{3/2}[\zeta(3/2)-\zeta(5/2)\lambda]  \sim \lambda^{3/2}>0$ is the free energy without considering quantum gravity effects at the non-relativistic limit,  $\Delta \tilde f^{NR}(\tilde n,\sigma)\sim -\sigma^2\tilde n^{5/3}< 0$ is the correct term when considering quantum gravity effects. The relative shift is $\Delta\tilde f/\tilde f^{NR}_0\sim -\Delta\lambda/\lambda$. 
At the ultra-relativistic limit, the Helmholtz free energy
\begin{eqnarray}
\tilde{f}&\simeq&\tilde{n} +\frac{\lambda^4}{\pi^2}  \int_{0^+}^\infty  {x^2}{(1-3\kappa^2 x^2)} \ln{\left[1-e^{-x}\right]}dx \nonumber \\
&\simeq& \left(\frac 13\lambda^2-\frac{\pi^2}{45}\lambda^4\right)+\frac {72\pi^4}{35}\sigma^2\tilde n^3\nonumber\\
&\equiv & \tilde f^{UR}_0+\Delta \tilde f^{UR}(\tilde n,\sigma),\label{kappad1freeenr2}
\end{eqnarray}
where $\tilde f^{UR}_0 =\frac 13\lambda^2-\frac{\pi^2}{45}\lambda^4\sim -\lambda^4<0$ is the free energy without considering quantum gravity effects at the ultra-relativistic limit,  $\Delta \tilde f^{UR}(\tilde n,\sigma)\sim\sigma^2\tilde n^3 > 0$. Also, the relative shift is $\Delta\tilde f/\tilde f^{UR}_0\sim -\Delta\lambda/\lambda$. 

\begin{figure}[h]
\centering \includegraphics[height=5cm]{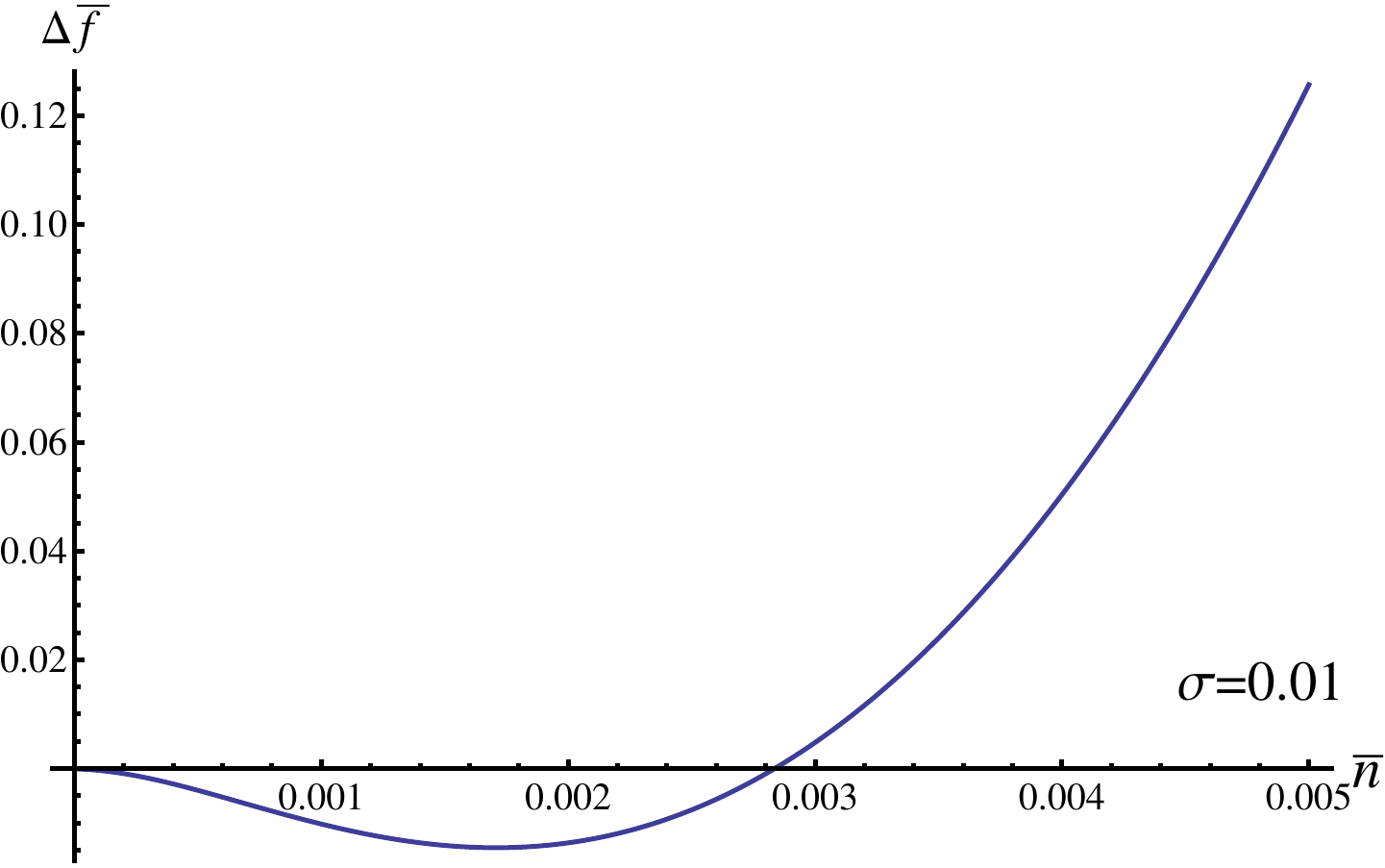}\caption {For a fixed $\sigma=0.01$,
$\Delta \bar f$ versus $\bar n$. When $\bar n$ is small enough, $\Delta \bar f<0$.
$\Delta \bar f$ has a minimum around $\bar n_{min}=0.0018$. When $\bar n$ is sufficiently large, $\Delta \bar f>0$.}.\label{fbarVSnsigma001}
\end{figure}

\begin{figure}[h]
\centering \includegraphics[height=5cm]{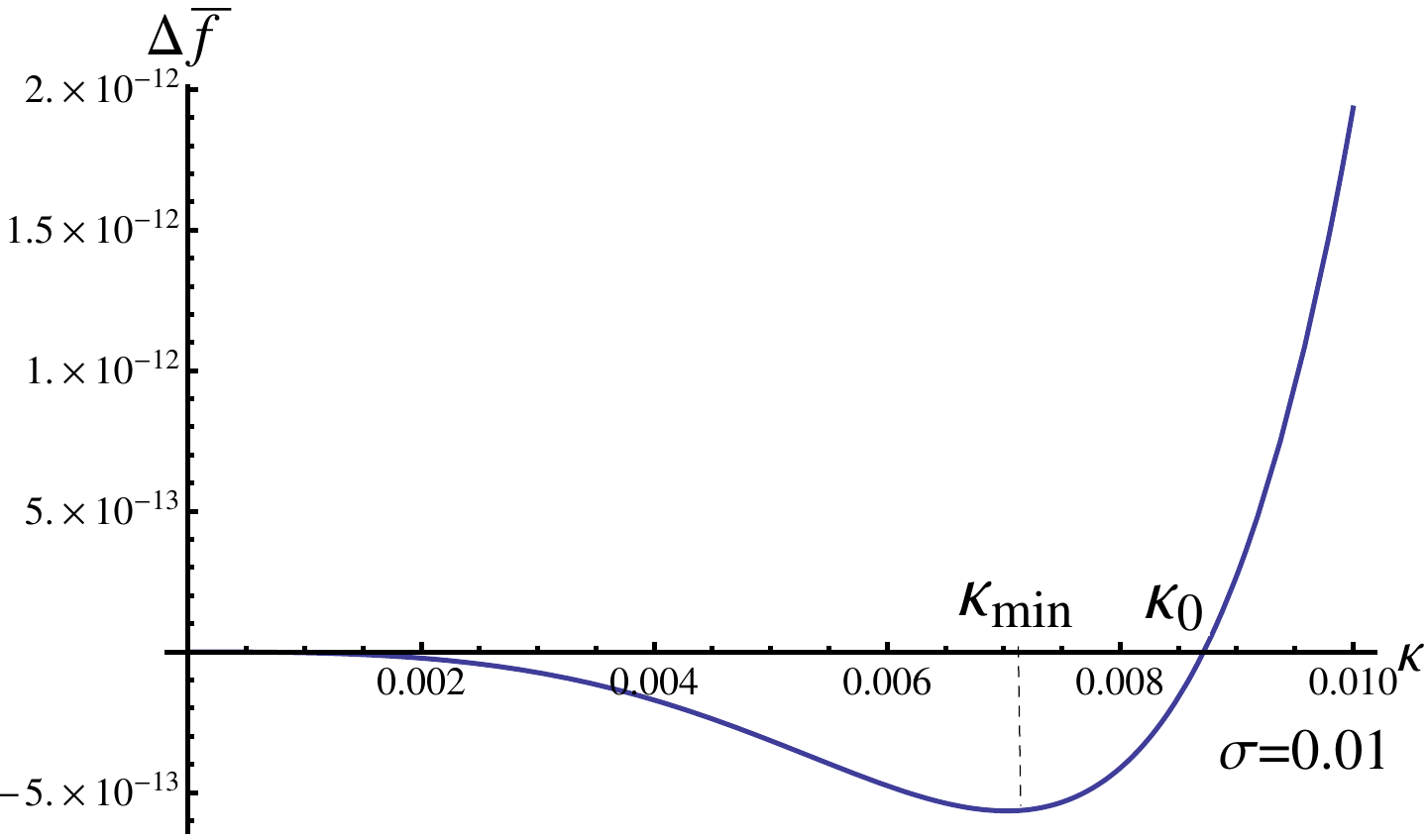} \caption {For a fixed $\sigma=0.01$,
$\Delta \bar f$ versus $\kappa=T_c/T_\textrm{H}$.
$\Delta \bar f$  has a minimum around $\kappa_{\textrm{min}}=0.007$. The stable-unstable transition temperature $\kappa_0\approx 0.0087$.}\label{deltafbarVSkappa001}
\end{figure}

\begin{figure}[h]
\centering \includegraphics[height=5cm]{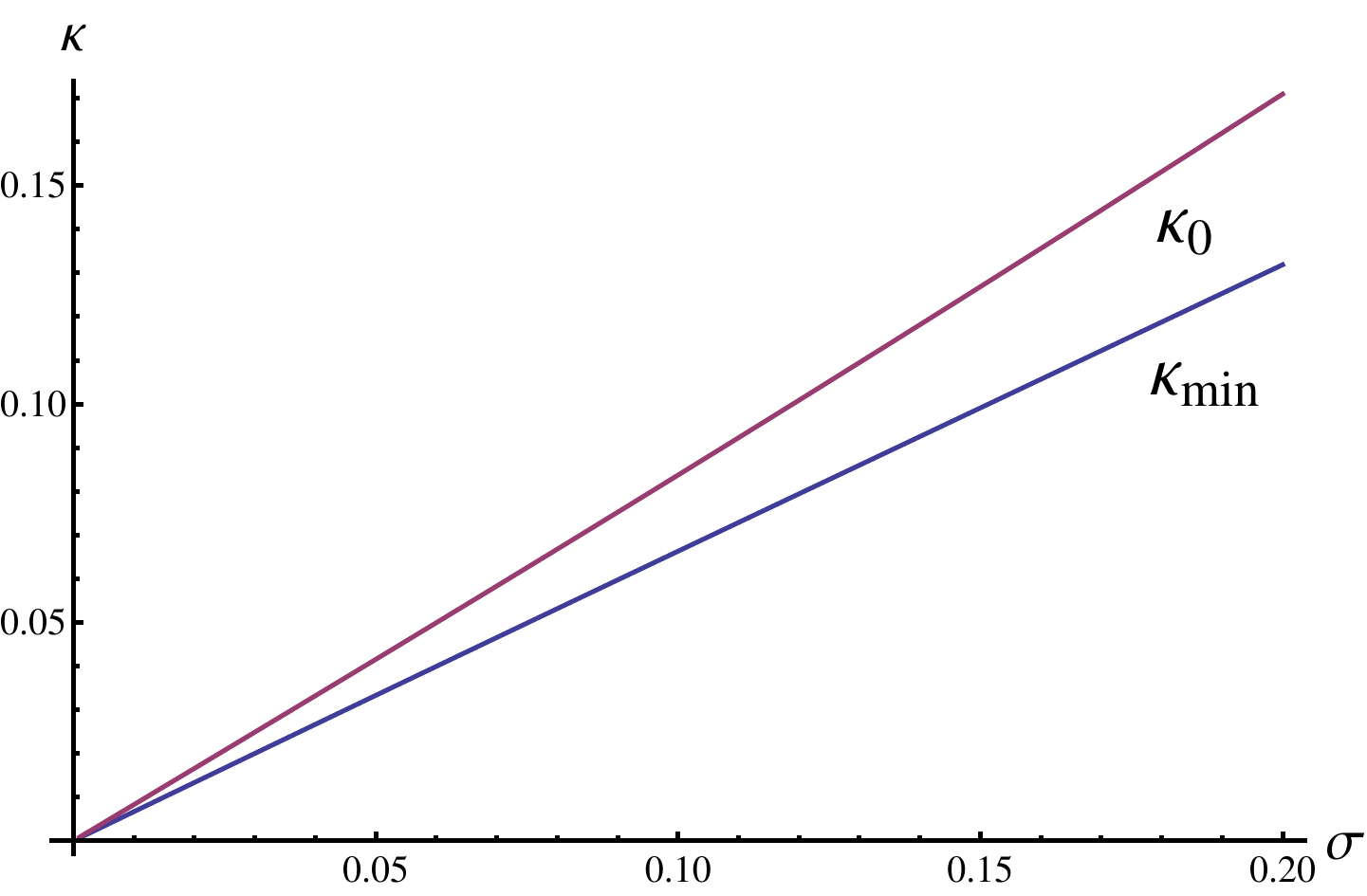} \caption {$\kappa_{\textrm{min}}$ corresponding to the minimum value of $\Delta \bar f$  and the stability-instability transition critical temperature $\kappa_0$ versus different values of $\sigma=\Delta_{\textrm{min}}/\Delta_C$.}\label{kappaminvssigma}
\end{figure}

For a fixed $\sigma$, we rescale the number density and the Helmholtz free energy as
$$
\bar{n}\equiv \sigma^3 \tilde n=(\Delta_{min}/\bar d)^3,\;\bar f\equiv\sigma^4\tilde f= f\Delta_{min}^3/E_\textrm{H}.
$$
Then in Eq. (\ref{Hfreeenergy1}), we write $\Delta \bar f=\bar f (\bar n,\sigma)-\bar f_0$, where $\bar f_0$ is the free energy without considering quantum gravity effects. FIG. \ref{fbarVSnsigma001} shows $\Delta \bar f$ versus $\bar n$ for a fixed $\sigma=0.01$. When $0<\bar n<\bar n_0\approx 0.0028$, $\Delta \bar f<0$. When $\bar n>\bar n_0\approx 0.0028$, $\Delta \bar f>0$. These results are consistent with previous analytical results in (\ref{kappax1freeenr1}) and (\ref{kappad1freeenr2}). Note that in FIG. \ref{fbarVSnsigma001}, $\bar f$ and $\bar f_0$ correspond to different critical temperatures respectively for the same density $\bar n$. Since $\lambda (\tilde n,\sigma)$ is a monotonous function of $\tilde{n}$, we rewrite $\Delta \bar f=\bar f (\kappa,\sigma)-\bar f_0$. FIG. \ref{deltafbarVSkappa001} shows $\Delta \bar f$ versus $\kappa=T_c/T_\textrm{H}$ for a fixed $\sigma=0.01$.
There exists a stable-unstable transition temperature $\kappa_0$. When $0<\kappa<\kappa_0\approx 0.0087$, $\Delta \bar f<0$, quantum gravity effects lower the Helmholtz free energy and then make BEC stabler while when $\kappa>\kappa_0$, $\Delta \bar f>0$, quantum gravity effects lift the Helmholtz free energy and tend to make BEC unstabler.
Furthermore, $\Delta \bar f$ has a minimum around $\kappa_{\textrm{min}}\approx 0.007$. Obviously, $\kappa_0$ and $\kappa_{\textrm{min}}$ depend on $\sigma$. FIG. \ref{kappaminvssigma} depicts $\kappa_{\textrm{min}}$ and $\kappa_0$ versus different values of $\sigma$.


In summary, we discussed the relativistic
ideal Bose gas at the critical temperature of Bose-Einstein condensation by a simple effective quantum gravity model. When the critical temperature $T_c\ll E_\textrm{H}/k_\textrm{B}$, we obtained the analytic results of the shift in the critical temperature and the shift of the Helmholtz free energy due to quantum gravity at the non-relativistic limit and the ultra-relativistic limit. The exact numerical results of these shifts are given. 
By measuring the shift of the critical temperature, we could measure and constrain the deformation parameter $\beta_0$ accounting for quantum gravity effects.
It is shown that at sufficiently low temperatures or densities, the Helmholtz free energy with quantum gravity effects is lower, thus implying that omitting quantum gravitational effects may lead to a metastable state. However, at sufficiently high temperatures or densities, the Helmholtz free energy with quantum gravity effects is higher, thus implying that quantum gravity effects tend to make the BEC unstaber. Numerical methods show that the stable-unstable transition temperature $\kappa_0$ depends on the deformation parameter $\beta_0$ and boson mass $m$, that is, $\sigma={\Delta_{min}}/{\Delta_C}=\beta_0m/M_p$. Our results are dependent on the GUP model. More refined model of quantum gravity may improve outcomes for details.  It would be of importance in the future work to explore quantum gravity effects on
the critical behavior of the relativistic ideal Bose gas and apply to boson compact stars and BECs in the early universe.

\textbf{Acknowledgements}

We are grateful to Xiao Liu for useful discussions. This work is supported by "the Fundamental Research Funds for the Central Universities", and in part by NSFC (Grant No. 11205125)



\begin{thebibliography}{1}

\small


%
\bibitem{Amelino-Camelia2002MPLA}
G. Amelino-Camelia, Mod. Phys. Lett \textbf{A 17} (2002) 899.
%
\bibitem{Amelino-Camelia2013LRR}
G. Amelino-Camelia, 
Living Rev. Rel. \textbf{16} (2013) 5. 
\bibitem{Arkani-Hamed1999PRD}
N. Arkani-Hamed, S. Dimopoulos and G. Dvali, Phys. Rev. \textbf{D 59} (1999) 086004.
%
\bibitem{Amelino-Camelia1998nature393}
G. Amelino-Camelia, J. Ellis, N. E. Mavromatos, D. V. Nanopoulos and S. Sarkar, Nature \textbf{393} (1998) 763.
%
\bibitem{Biler1999PRL83}
S. D. Biller et al., Phys. Rev. Lett. \textbf{83} (1999) 2108.
%
\bibitem{Kifune1999APJL}
T. Kifune, Astrophys. J. Lett. \textbf{518} (1999) L21.
%
\bibitem{Protheroe2000PLB}
R. J. Protheroe and H. Meyer, Phys. Lett. \textbf{B 493} (2000) 1.
%
\bibitem{Amelino-Camelia2001PRD}
G. Amelino-Camelia and T. Piran, Phys. Rev. \textbf{D 64} (2001) 036005.
%
\bibitem{TJacobson2002PRD}
T. Jacobson, S. Liberati and D. Mattingly, Phys. Rev. \textbf{D 66} (2002) 081302.
%
\bibitem{Amelino-Camelia2002PLB}
G. Amelino-Camelia, Phys. Lett. \textbf{B 528} (2002) 181.
%
\bibitem{Alfaro2000PRL}
J. Alfaro, H. A. Morales-Tecotl and L. F. Urrutial, Phys. Rev. Lett. \textbf{84} (2000) 2318.
%
\bibitem{Jacob2007NaturePhysics}
U. Jacob and T. Piran, 
Nature Physics \textbf{3} (2007) 87.
%
\bibitem{Amelino-Camelia1999Nature398}
G. Amelino-Camelia, Nature \textbf{398} (1999) 216.
%
\bibitem{YJNg2000FP}
Y. Jack Ng and H. van Dam, 
Foundations of Physics \textbf{30} (2000) 795.
%
\bibitem{Amelino-Camelia2000PRD62}
G. Amelino-Camelia, Phys. Rev. \textbf{D 62} (2000) 024015.
%
\bibitem{Amelino-Camelia2004CQG}
G. Amelino-Camelia and C. L$\ddot{\textrm{a}}$mmerzahl,
Class. Quant. Grav. \textbf{21} (2004) 899.
\bibitem{Hogan2012PRD}
C. J. Hogan, 
Phys. Rev. \textbf{D 85} (2012) 064007.
%
%
\bibitem{Amelino-Camelia2009PRL}
G. Amelino-Camelia, C. Lammerzahl, F. Mercati and G. M. Tino, Phys. Rev. Lett. \textbf{103} (2009) 171302.
%
\bibitem{Mercati2010CQG}
F. Mercati \textit{et al}, Class. Quan. Grav. \textbf{27} (2010) 215003.
\bibitem{BrisceseGrether2012EPL}
F. Briscese, M. Grether and M. de Llano, 
Europhys. Lett. \textbf{98}  (2012) 60001.
%
\bibitem{Briscese2012PLB}
F. Briscese, Phys. Lett. \textbf{B 718} (2012) 214.
%
\bibitem{Castellanos2013EPL}
E. Castellanos, Eur. Phys. Lett. 103 (2013) 40004.
%
\bibitem{Castellanos2012MPLA}
E. Castellanos and C. Laemmerzahl, Mod. Phys. Lett. \textbf{A} 27 (2012) 1250181.
%
\bibitem{Castellanos2014PLB}
E. Castellanos and C. Laemmerzahl, Phys. Lett. \textbf{B} 731 (2014) 1.
%

%
\bibitem{Garay1995IJMPA}
L. J. Garay, Int. J. Mod. Phys. \textbf{A 10} (1995) 145.
%
\bibitem{Scardigli1999PLB}
F. Scardigli, Phys. Lett. \textbf{B 452} (1999) 39.
%
\bibitem{Kempf1995PRD}
A. Kempf, G. Mangano and R. B. Mann, Phys. Rev. \textbf{D 52 }(1995) 1108.
%
\bibitem{LNCMinic2002PRD}
L. N. Chang, D. Minic, N. Okamura and T. Takeuchi,
Phys. Rev. \textbf{D 65}  (2002) 125028.
%
%
\bibitem{Fityo2008PLA}
T. V. Fityo, Phys. Lett. \textbf{A} 372 (2008) 5872.
%
%
\bibitem{Xiumzhang2010JHEP}
Peng Wang , Haitang Yang and Xiuming Zhang,  
JHEP \textbf{08} (2010) 043.
265.
%
\bibitem{Xiumzhang2012PLB}
Peng Wang , Haitang Yang and Xiuming Zhang,  
Phys. Lett. \textbf{B 718} (2012) 265.
%
\bibitem{BrauBuisseret2006PRD}
F. Brau, F. Buisseret, Phys. Rev. \textbf{D 74 }(2006) 036002.
%
\bibitem{DasVagenas2008PRL}
S. Das and E. C. Vagenas, 
Phys. Rev. Lett. \textbf{101}  (2008) 221301.
%
\bibitem{LAULopez2009JCAP}
L. A. Ure$\tilde{\textrm{n}}$a-L$\acute{\textrm{o}}$pez, JCAP\textbf{ 01} (2009) 014.
%
\bibitem{Grether2008IJMPB}
 M. Grether, M. De Llano, S. Ram$\acute{\textrm{i}}$rez and O. Rojo, Int. J. Mod. Phys. \textbf{B 22} (2008) 4367.
%



\end{thebibliography}
\end{document}